\documentclass[prl,twocolumn,epsf,psfig]{revtex4}
\usepackage{graphicx}

\begin{document}

\title{Fermionic superfluid properties in a one-dimensional optical lattice }

\author{Theja N. De Silva}
\affiliation{Department of Physics, Applied Physics and Astronomy,
The State University of New York at Binghamton, Binghamton, New York
13902, USA.}
\begin{abstract}
We discuss various superfluid properties of a two-component Fermi
system in the presence of a tight one-dimensional periodic potential
in a three-dimensional system. We use a zero temperature mean field
theory and derive analytical expressions for the Josephson current,
the sound velocity and the center of mass oscillations in the
BCS-Bose Einstein condensation crossover region.
\end{abstract}

\maketitle

\section{I. Introduction}

Recent experiments with trapped alkaline Fermi gases have shown
promising directions toward understanding the many body physics in
various disciplines. These systems have became a focuss of attention
due to their easy controllability. For example, the controllability
of interaction strength using a magnetically tuned Feshbach
resonance~\cite{fb} and of the spatial structure using counter
propagating laser beams allow one to study a wide range of quantum
mechanical phenomena. One such phenomena is a dissipationless flow
at low temperatures. The focus of this paper is the analytical study
of dissipationless transport properties of a stack of weakly-coupled
two-dimensional (2D) Fermi systems. Such a system can be created by
applying a relatively strong one-dimensional (1D) optical lattice
potential to an ordinary three-dimensional system. One natural
analogy of this system is the weakly coupled copper-oxide planes of
high temperature superconductors where Bose condensed paired
electrons flow dissipationlessly in the crystal lattice.

In this paper, we derive analytical expressions for the energy and
the density in the limit of weakly coupled layers using a zero
temperature mean field theory. Using the energy and the density
expressions together with gap equation, we then derive analytical
expressions for various dynamical quantities in the Bardeen, Cooper
and Schrieffer-Bose Einstein condensation (BCS-BEC) crossover
region. A similar discussion of tunneling dynamics using a path
integral approach can be found in ref.~\cite{devreese1}. A
hydrodynamic approach in the weakly interacting BCS limit and a
Bogoliubov-de Gennes approach in the unitarity limit can be found in
ref.~\cite{stringari1} and ref.~\cite{stringari2} respectively.
First, we calculate the superfluid oscillation (Bloch oscillation)
amplitudes due to the linear force on atoms in the presence of Bloch
bands and find that the amplitudes are larger for weak periodic
potentials and for weak attractive interactions. Second, we
calculate the sound velocity along the direction of 1D lattice
potential and find that the sound velocity strongly depends on the
lattice height but weakly depends on the interaction for tight
periodic potentials. Finally, we calculate the dipole oscillation
frequencies in the BCS-BEC crossover region in the presence of a
harmonic external trapping potential. In the absence of a periodic
potential, dipole oscillations are undamped and the oscillation
frequencies are independent of the two-body interaction (generalized
Kohn's theorem~\cite{kohn}). However, in the presence of a 1D
periodic potential, we find that the dipole oscillations are damped
and strongly depend on the two-body interactions. Such damping for a
1D Bose gas and a non-interacting Fermi gas have already been
observed~\cite{damp}.

The paper is organized as follows. In section II, we present the
theoretical formulation. In section III, we present the results and
discussion followed by the conclusion in section IV.

\section{II. Formalism}

We consider an interacting two-component Fermi atomic gas trapped in
one dimensional optical lattice created by standing laser waves. The
1D optical potential modulated along z-axis has the form $V =
sE_R\sin^2 (2\pi z/\lambda)$. Here $\lambda$ is the wavelength of
the of the laser beam, $E_R = \hbar^2(2\pi/\lambda)^2/(2M)$ is the
recoil energy and $s$ is a dimensionless parameter which accounts
the intensity of the laser beam. The periodicity of the 1D optical
lattice along z-direction is $d = \lambda/2$. When the laser power
is large, the atomic system forms a stack of weakly coupled 2D
planes. We take the model Hamiltonian of the system as $H =
\sum_jH_j$, where $j$ is the layer index.

\begin{eqnarray}\label{model1}
H_j = \int
d^2\vec{r}\biggr\{\sum_{\sigma}\psi^{\dagger}_{j\sigma}(r)[-\frac{\nabla^2_{2D}}{2M}+V_{ho}(j)-\mu_{\sigma}]\psi_{j\sigma}(r)
\\ \nonumber + t
\sum_{\sigma}[\psi^{\dagger}_{j\sigma}(r)\psi_{j+1\sigma}(r)+hc] \\
\nonumber
+U_{2D}\psi^{\dagger}_{j\uparrow}(r)\psi^{\dagger}_{j\downarrow}(r)\psi_{j\downarrow}(r)\psi_{j\uparrow}(r)\biggr\}
\end{eqnarray}

\noindent where $V_{ho} = 1/2M\omega_{\perp}^2r^2
+1/2M\omega_z^2a^2j^2$ is the external trapping potential with $r^2
= x^2+y^2$, $\nabla_{2D}$ is the 2D gradient operator and $U_{2D}$
is the 2D interaction strength. The operator
$\psi^{\dagger}_{j\sigma}(r)$ creates a fermion of mass $M$ in in
$j$th plane with spin $\sigma = \uparrow, \downarrow$ at position $r
= (x, y)$. We consider a tight 1D lattice where interlayer tunneling
energy $t$ is small and consider the following expression proposed
in ref.~\cite{hopping} and ref.~\cite{jaksch} for the weakly coupled
layers.

\begin{figure}
\includegraphics[width=\columnwidth]{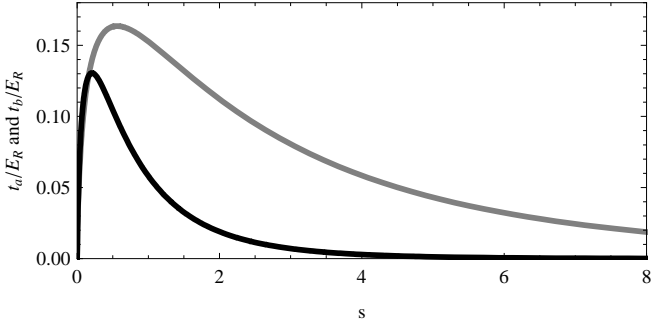}
\caption{Tunneling energies as function of laser intensity. Gray
line: Tunneling energy ($t_{b}$) proposed in ref~\cite{jaksch} for
3D optical potentials. Black line: Tunneling energy ($t_a$) proposed
in ref.~\cite{hopping} for 1D optical potentials. The dimensionless
parameter $s$ accounts the laser intensity.} \label{hop}
\end{figure}

\noindent In terms of dimensionless parameters, the tunneling energy
for a 1D lattice proposed in ref.~\cite{hopping} is $t_a/E_R = \pi s
(\pi^2/4-1)\exp[-\pi^2\sqrt{\pi s}/4]$. This contrasts with the
isotropic three-dimensional (3D) lattice tunneling parameter
$t_{b}/E_R = (2s^{3/4}/\sqrt{\pi})\exp[-2\sqrt{s}]$ given in
ref.~\cite{jaksch}. Both of these are derived using a harmonic
approximation around the minima of the optical lattice potential.
Notice that these ratios can be varied by changing the laser
intensity. As depth of the optical potential increases, the atomic
wave function becomes more and more localized and the tunneling
amplitude decreases. For comparison, these tunneling amplitudes are
shown in FIG.~(\ref{hop}) as a function of the laser intensity. For
asymptotically deep lattices and any separable potential, the
tunneling energy should be independent of the dimensionality of the
lattice. The asymptotically difference between $t_a$ and $t_b$ seen
in FIG.~(\ref{hop}) may be due to the neglecting of the
non-orthogonality of the adjacent lattice Wannier functions in
ref.~\cite{hopping}. In this paper, we use the expression $t = t_b$
for all our calculation. In the limit $t \rightarrow 0$, the system
is decoupled planes of Fermi atoms. We use mean field theory to
decouple the interaction term writing
$U_{2D}\psi^{\dagger}_{j\uparrow}(r)\psi^{\dagger}_{j\downarrow}(r)\psi_{j\downarrow}(r)\psi_{j\uparrow}(r)
= [\Delta^{\dagger}_j(r)
\psi_{j\downarrow}(r)\psi_{j\uparrow}(r)+h.c]-|\Delta_j(r)|^2/U_{2D}$.
The Fourier transform of the Hamiltonian gives,

\begin{eqnarray}\label{model2}
H= \sum_{k,\sigma m}(\epsilon_k - \mu_{\sigma})a^{\dagger}_{m\sigma
}(k) a_{m\sigma}(k) \\ \nonumber+ t \sum_{k,\sigma
m}[a^{\dagger}_{m+1\sigma}(k) a_{m\sigma}(k) + h.c] \\ \nonumber
+\sum_{km}[\Delta_ma^{\dagger}_{m\uparrow}(k)
a^{\dagger}_{m\downarrow}(-k)+h.c]-\sum_{mj}\frac{|\Delta_m|^2}{U_{2D}}
\end{eqnarray}

\noindent where $\epsilon_k = \hbar^2k^2/(2M)$ with $k^2 =
k_x^2+k_y^2$. Notice that we have absorbed the external trapping
potential into the chemical potential. One can treat this local
chemical potential with local density approximation. In this paper,
we take the external trapping potential into account when we
calculate the dipole oscillations, otherwise we treat the system as
homogenous. In order to take into account superfluid flow along
z-direction, we take superfluid order parameter in the form
$\Delta_m = \Delta \exp[im\phi]$. The periodicity along z-direction
allows us to write the Fermi operators,

\begin{eqnarray}\label{ft}
a_{m\uparrow}(k) = \sum_{k_z}\exp[ik_zmd+im\phi/2]c_{\uparrow}(k)
\nonumber \\
a_{m\downarrow}(-k) =
\sum_{k_z}\exp[-ik_zmd+im\phi/2]c_{\downarrow}(-k)
\end{eqnarray}

Transforming the Hamiltonian given in Eq. (\ref{model2}) using above
transformation followed by the usual Bogoliubov transformation, the
Hamiltonian per plane can be expressed as~\cite{tanaka}

\begin{eqnarray}\label{ft}
H/N =
\sum_{k,k_z,\sigma}[\sqrt{\bar{\epsilon}_k-\mu)^2+\Delta^2}-2t\sin
(k_zd)\sin(\phi/2)]\\
\nonumber
(\alpha^{\dagger}_{k\sigma}\alpha_{k\sigma}+hc)-\sum_{k,k_z}[\sqrt{\bar{\epsilon}_k-\mu)^2+\Delta^2}-\bar{\epsilon}_k+\mu]-\frac{\Delta^2}{U_{2D}}
\end{eqnarray}

\noindent where $N$ is the number of planes and $\bar{\epsilon}_k =
\epsilon_k + 2t \cos (k_zd) \cos(\phi/2)$. Notice that the quasi
particle energy depends on the phase of the superfluid order
parameter. In the normal phase, the phase factor $\phi$ can have
multiple values of $\pi$~\cite{note}. The grand potential of the
system $\Omega = (-1/\beta) \ln[Z_G]$ with $Z_G = Tr \exp[-\beta H]$
at zero temperature is

\begin{eqnarray}\label{grand pot1}
\Omega = \sum_{k_z}\int \frac{d^2k}{(2
\pi)^2}\biggr[\bar{\epsilon}_k-\mu-\sqrt{(\bar{\epsilon}_k-\mu)^2+\Delta^2}\biggr]-\frac{\Delta^2}{U_{2D}}
\end{eqnarray}

\noindent The 2D contact interaction $U_{2D}$ is related to the
bound state energy $E_B$ as~\cite{u2d}

\begin{eqnarray}\label{u2d}
\frac{1}{U_{2D}} = - \int \frac{d^2k}{(2\pi)^2}
\frac{1}{\hbar^2k^2/M+E_B}
\end{eqnarray}

\noindent The bound state state energy $E_B = (C\hbar \omega_L/\pi)
\exp[\sqrt{2\pi}l_L/a_s]$, where $a_s$ is the 3D s-wave scattering
length and $C \approx 0.915$. Notice that the bound state energy
depends not only on the scattering length but also on the lattice
potential. Performing the momentum integrals, the grand potential is

\begin{eqnarray}\label{grand pot2}
\Omega =
\frac{m}{2\pi\hbar^2}\biggr\{\biggr(-\frac{\mu^2}{2}-\frac{\Delta^2}{4}-\frac{\mu}{2}\sqrt{\mu^2+\Delta^2}\biggr)\\
\nonumber -\biggr(1 +
\frac{3\Delta^2\mu+2\mu^3}{2(\mu^2+\Delta^2)^{3/2}}\biggr)\cos^2(\phi/2)t^2
\\ \nonumber + \frac{15 \Delta^4 \mu}{8 (\mu^2+\Delta^2)^{7/2}}\cos^4(\phi/2)t^4
+ {\cal O}(t^6)\biggr\}
\end{eqnarray}

\noindent Then the gap equation and the number density are
calculated by using the equations, $\partial \Omega/\partial \Delta
= 0$ and $n = -\partial \Omega/\partial \mu$, respectively.

\begin{eqnarray}\label{gap}
\ln \biggr[\frac{E_B}{-\mu+\sqrt{\mu^2+\Delta^2}}\biggr]-\frac{\mu
\cos^2 (\phi/2)}{(\mu^2+\Delta^2)^{3/2}}t^2
\\ \nonumber +\frac{9\Delta^2\mu-6\mu^3}{4(\mu^2+\Delta^2)^{7/2}}\cos^4(\phi/2)t^4+
{\cal O}(t^6)=0
\end{eqnarray}

\begin{eqnarray}\label{density}
n =
\frac{m}{2\pi\hbar^2}\biggr\{\biggr(\mu+\sqrt{\mu^2+\Delta^2}\biggr)+
\frac{\Delta^2\cos^2(\phi/2)}{(\mu^2+\Delta^2)^{3/2}}t^2
\\ \nonumber -\frac{3\Delta^2(\Delta^2-4\mu^2)}{\mu^2+\Delta^2)^{7/2}}\cos^4(\phi/2)t^4
+ {\cal O}(t^6)\biggr\}
\end{eqnarray}

\begin{figure}
\includegraphics[width=\columnwidth]{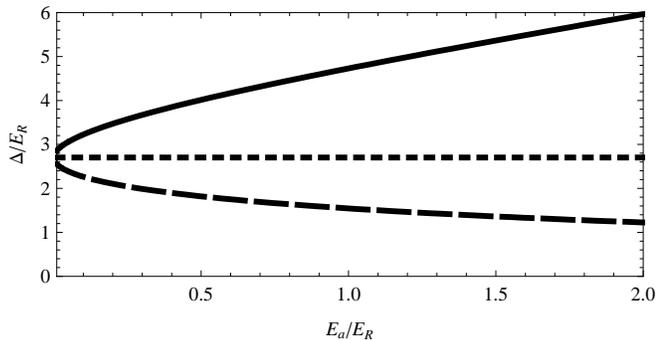}
\caption{Superfluid order parameter $\Delta$ as a function of $E_a =
\hbar^2/(2Ma_s^2)$ for $a_s < 0$ (long dashed line), $a_s
\rightarrow \infty$ (short dashed line) and $a_s >0$ (solid line) at
fixed density $n = 2.5 m/(2\pi \hbar^2)$ and $s =8$. } \label{del}
\end{figure}

\section{III. results and discussion}

All our derivations for physical quantities will be calculated from
Eq.~(\ref{grand pot2}), Eq.~(\ref{gap}) and Eq.~(\ref{density}). For
example, the solutions of Eq. (\ref{gap}) for various parameters are
shown in FIG.~(\ref{del}). For the case of decoupled layers (in the
limit $t = 0$), Eq.~(\ref{gap}) gives $\Delta =
\sqrt{E_B(E_B+2\mu)}$. This results the grand potential $\Omega =
-[m/(8\pi\hbar^2)](E_B+2\mu)^2$ and the number density $n =
m/(2\pi\hbar^2)](E_B+2\mu)$.

In the following subsections we calculate the Josephson current, the
sound velocity and the dipole oscillation frequencies. First we
calculate the oscillatory superfluid velocity along the direction of
optical lattice through the Josephson current in the absence of
harmonic potential. Next, we consider the propagation of sound along
z-direction and calculate the sound velocity in a homogenous system.
Later we include the external harmonic potential to investigate the
dipole oscillations of the atomic cloud.

\subsection{Josephson current and Bloch oscillations}

For the solid-state electronic systems, the size of the Cooper pair
is usually larger than the lattice spacing and one can treat the
system as homogenous when calculating physical quantities such as
the supercurrent. However, the periodicity of the cold atomic system
is adjustable. As a result one can expect new phenomena on optical
lattices with Fermi superfluid. For example, oscillating fermionic
superfluidity which is absent in the bulk solid can be observed with
the periodic potentials. These oscillations may be used as a probe
of BCS-BEC crossover and of the normal to superfluid transitions. We
calculate the particle supercurrent along z-direction using $j(\phi)
= (1/\hbar)
\partial \Omega/\partial \phi$.

\begin{eqnarray}\label{current}
j(\phi) = \frac{m}{2\pi\hbar^2}\frac{\sin\phi}{\hbar}
\biggr\{\biggr(\frac{1}{2}+\frac{3\Delta^2\mu+2\mu^3}{4
(\mu^2+\Delta^2)^{3/2}}\biggr)t^2
\\ \nonumber -\frac{15\Delta^4\mu}{8(\mu^2+\Delta^2)^{7/2}}\cos^2(\phi/2)t^4
+{\cal O}(t^6)\biggr\}
\end{eqnarray}

\begin{figure}
\includegraphics[width=\columnwidth]{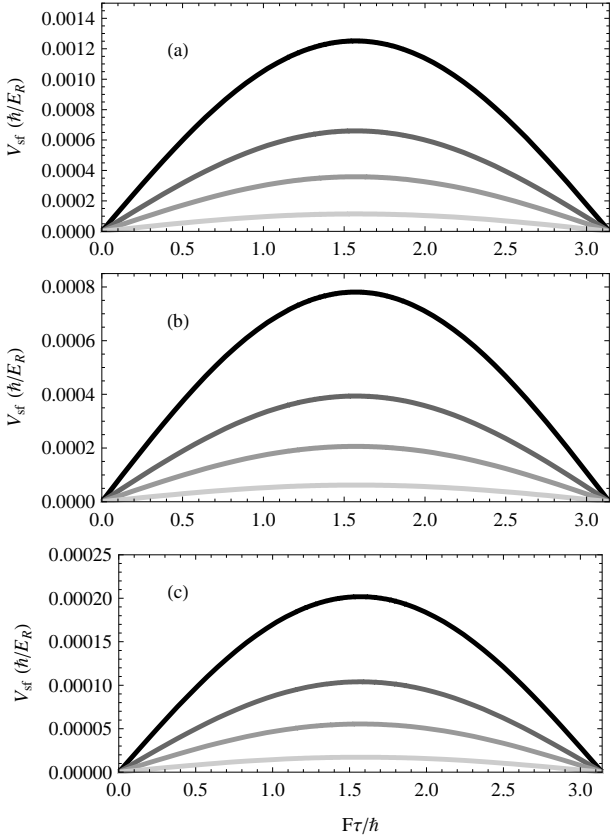}
\caption{Block oscillations in (a). BCS regime, (b). Unitarity
regime, and (c). BEC regime, where we fixed the scattering length to
be $a_s = -\sqrt{\hbar^2/(2mE_R)}$, $a_s \rightarrow \infty$, and
$a_s = +\sqrt{\hbar^2/(2mE_R)}$ respectively. The parameter $s$
varies as $4$ (black), $5$ (dark gray) , $6$ (light gray) and, $8$
(lighter gray). For all the calculation, we use the density $n = 2.5
m/(2\pi \hbar^2)$.} \label{sfv}
\end{figure}

In early quantum theory of electrical conductivity, Bloch and
Zener~\cite{block} predicted that when a static electric field is
applied to a crystalline electronic system, instead of uniform
motion one would naively expect, electron in the crystal should
oscillate. However these so-called Bloch oscillations have never
been observed in the bulk crystals. The reason is that the period of
Bloch oscillations is much larger than the electron scattering rate.
As the cold atomic superfluid in optical lattices are clean and
controllable, these systems have shown to be well suited  for the
observation of Bloch oscillations. In fact, the oscillations are
already observed in one dimensional optical potentials~\cite{boex}.

We use superfluid velocity along the z-direction to get some
understanding about the amplitude of the Bloch oscillations.
Consider an atom is in the Bloch state $|l, k_0>$, where $l$ is the
discrete band index and $k_0$ is the continuous quasi momentum. If
the atom feels a constant force $F$ (In optical lattices, one can
mimic a constant force on atoms by introducing a tunable frequency
difference between two counter propagating laser beams), the above
Bloch state evolve to a another state $|l, k(\tau)>$, where $k(\tau)
= k_0 + F\tau/\hbar$  is the quasi momentum at time $\tau$. We
assumed that the force is weak enough and the 1D lattice is tight
enough not to induce the inter band transitions. Therefore setting
$\phi = F\tau/\hbar$ in Eq.~(\ref{current}), we can calculate
current due to the induced Bloch oscillations. Then we calculate the
superfluid velocity using $v_{SF}(F\tau/\hbar) = j(F\tau/\hbar)/n$.

\begin{eqnarray}\label{current}
v_{SF} = \frac{\sin(F\tau/\hbar)}{\hbar} \biggr\{\frac{2\Delta^2 +
\mu^2 + \mu \sqrt{\mu^2+\Delta^2}}{4 (\mu^2+\Delta^2)^{3/2}}t^2 \\
\nonumber -
\frac{\mu(4\mu^2+21\Delta^2)\sqrt{\Delta^2+\mu^2}+4(\Delta^4+\mu^4)+23\Delta^2\mu^2}{8
(\Delta^2+\mu^2)^{7/2}(\mu+\sqrt{\Delta^2+\mu^2})^2} \times \\
\nonumber \Delta^2\cos^2(F\tau/2\hbar)t^4 +{\cal O}(t^6)\biggr\}
\end{eqnarray}

\begin{figure}
\includegraphics[width=\columnwidth]{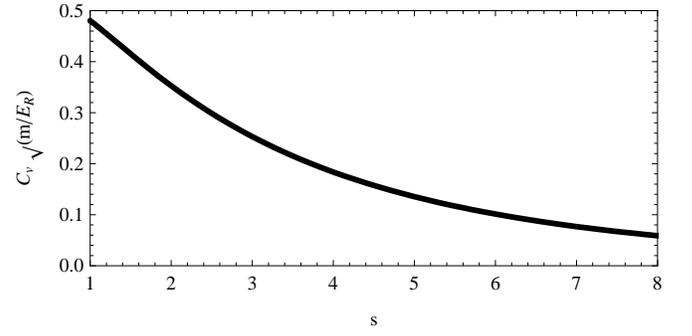}
\caption{Sound velocity along the direction of 1D lattice as a
function of dimensionless parameter $s$. For small tunneling
energies (tight 1D potentials), the sound velocity is almost the
same in the entire BCS-BEC crossover region. } \label{svelo}
\end{figure}

These superfluid velocities are shown in FIG.~(\ref{sfv}) for three
different regimes. It is clear that the Bloch oscillations are
stronger (the amplitude is larger) in the BCS regime where $\Delta$
is smaller and where the coherence length is large compared to the
inter layer distance. In other words, away from the BCS regime the
effective mass of the Cooper pair becomes larger (due to the strong
interaction) and localizes in each plane. In order to observe these
Bloch oscillations, superfluid velocity should not be exceeded the
critical velocity. This Landau criterion for superfluidity is
satisfied if the Cooper pair size is smaller than the lattice
periodicity.

\subsection{Propagation of sound}

We consider a homogenous system and the propagation of the sound
along z-direction. The form of the phonon dispersion $\omega = c_zq$
gives the sound velocity along z-direction $c_z =
\sqrt{(n/\tilde{M}) (\partial \mu/\partial n)}$, where the effective
mass of the atoms $\tilde{M}$ is given by $\tilde{M}^{-1} =\langle
\partial^2E_t/\partial k^2 \rangle$, where $E_t = 2t[1-\cos(k_zd/\hbar)]$. Using $\langle
\partial^2E_t/\partial k^2 \rangle = 2td^2/\hbar^2 \langle
\cos(k_zd/\hbar) \rangle =  (2td^2/\hbar^2((1/n) \sum_{k,k_z}n_k
\cos(k_zd/\hbar)$ with atom density $n = \sum_{k,k_z} n_k$ and
summing over the momentum states,

\begin{eqnarray}\label{efmass}
\frac{1}{\tilde{M}} =
\frac{2d^2}{\hbar^2}\biggr\{\frac{\cos(\phi/2)t^2}{\sqrt{\Delta^2+\mu^2}}
\\ \nonumber  - \frac{3\Delta^2
\mu+2\Delta^2\sqrt{\Delta^2+\mu^2}}{2(\Delta^2+\mu^2)^{5/2}(\mu+\sqrt{\Delta^2+\mu^2})}\times \\
\nonumber\cos^3(\phi/2)t^4 + {\cal O}(t^6)\biggr\}
\end{eqnarray}

\noindent Notice that the presence of periodic lattice is taken into
account through both the effective mass and the compressibility
$\partial \mu/\partial n$. And the ratio $c_\perp^2/c_z^2 =
\tilde{M}$, where $c_\perp$ is the sound velocity in transverse
direction. Calculating the compressibility using the number equation
(Eq.~(\ref{density}), the sound velocity is

\begin{eqnarray}\label{cz}
c_z =
\sqrt{\frac{2d^2}{\hbar^2}\cos(\phi/2)}\biggr\{t+\frac{3\Delta^2\mu
\cos^2(\phi/2)t^3}{4(\Delta^2+\mu^2)^2(\mu+\sqrt{\Delta^2+\mu^2})} \\
\nonumber + {\cal O}(t^5)\biggr\}
\end{eqnarray}

\begin{figure}
\includegraphics[width=\columnwidth]{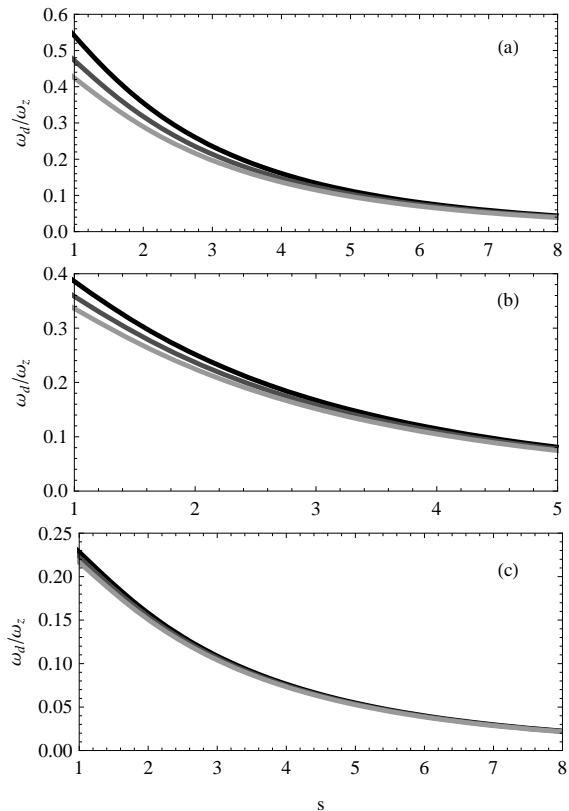}
\caption{Dipole oscillations in (a). BCS regime, (b). Unitarity
regime, and (c). BEC regime, where we fixed the scattering length to
be $a_s = -\sqrt{\hbar^2/(2mE_R)}$, $a_s \rightarrow \infty$, and
$a_s = +\sqrt{\hbar^2/(2mE_R)}$ respectively. The parameter $\mu$
varies as $0.5E_R$ (black), $0.75E_R$ (dark gray) and, $E_R$ (light
gray). } \label{dp}
\end{figure}

\noindent The sound velocity as a function of $s$ for $\phi = 0$ is
given in FIG.~(\ref{svelo}). As seen in Eq.~(\ref{cz}), the sound
velocity is independent of $\Delta$ and $\mu$ for the lowest order
in tunneling $t$. As a result, the sound velocity is almost same in
the entire BCS-BEC regime for a very tight 1D optical lattice. In
fact, this result holds for a non-interacting degenerate Fermi gas
in a very tight periodic potential. In the absence of a periodic
potential, the sound velocity is smaller than the Fermi velocity.
Therefore the sound cannot propagate in a degenerate non-interacting
Fermi gas. However, the periodic potential modifies the effective
mass of the atoms and the sound can propagate even in a
non-interacting Fermi gas.

\subsection{Dipole oscillations}
The Bloch Oscillations discussed in the previous subsection are due
to the atoms oscillations along the Bloch band under a linear force,
and the dipole oscillations  or the center of mass oscillations are
due to the confinement in the external trapping potential. Even
though the theoretical difference is clear, the probing of these two
types of oscillations may be very complicated. One possible way of
distinguishing these two types of oscillations is from observation
of the damping rate. If the momentum distribution is narrower than
the first Briloin zone, then the damping rate of the Bloch
oscillation will be zero.

The frequency of the dipole oscillations can be derived from the sum
rule approach. Defining the dipole operator $D = \sum_j^Nz_j$, the
energy weight moment $m_1 = (1/2)\langle[D,[H,D]]\rangle$ and the
inverse-energy weight moment $m_{-1} = N_0/(2M\omega_z^2)$ can be
calculated. Here $H$ is the Hamiltonian, $[A, B] = AB - BA$ is the
usual commutator, $N_0$ is the total number of atoms, and
$\langle..\rangle$ is average over the ground states. Then the upper
bound of the dipole oscillation frequency $\omega_d$ is calculated
from the expression~\cite{pita},

\begin{eqnarray}\label{dipole}
\hbar \omega_d =\sqrt{\frac{m_1}{m_{-1}}}
\end{eqnarray}

\noindent This gives the frequency of the center of mass oscillation
along the 1D lattice $\omega_d = \omega_z \sqrt{M/\tilde{M}}$.
Notice in the absence of optical lattice $\omega_d = \omega_z$ is
independent of the interaction.

The dipole oscillation frequency as a function of $s$ is given in
FIG.~(\ref{dp}) for $\phi = 0$. In the entire BCS-BEC region, the
dipole oscillations show significant reduction in the presence of 1D
lattice. This reduction is stronger in the BEC regime where $\Delta$
is larger. As evidence from the FIG.~(\ref{dp}), the oscillation
frequency shows a weak chemical potential (hence the total number)
dependence.

\section{IV. Conclusions}

We study some of the dynamical properties of a two-component Fermi
gas trapped in a parabolic plus 1D periodic potential. We use a zero
temperature mean field theory in the limit of tight 1D confinement.
We find that the Bloch oscillations become weaker as the interaction
strength changes from weakly attractive limit to weakly repulsive
limit through the unitarity regime. Furthermore, we find that the
sound velocity is independent of the interaction for very tight 1D
confinements. Finally, we find that the dipole oscillation
frequencies strongly depend on both the interaction and the 1D
periodic potential.

\section{V. Acknowledgements}
This work was supported by Binghamton University. We are very
grateful to Kaden Hazzard for useful discussions on the dimensional
dependence of the tunneling energies.


\begin{references}
\bibitem{note} By changing the quasi momentum of the atoms adiabatically (for example, using a linear force on the atoms), one can have arbitrary values for $\phi$ in the normal
phase.
\end{references}
\end{document}